\documentclass[11pt]{article}
\pdfoutput=1
\usepackage{amsmath}
\usepackage{amssymb}
\usepackage{graphicx,bbm,mathrsfs}
\usepackage{nicefrac}
\usepackage{bbm}
\usepackage{geometry}       
\usepackage{jheppub}     

\usepackage{dcolumn}   
\usepackage{bm}        
\usepackage{graphicx,mathrsfs}
\usepackage{nicefrac}
\usepackage{multirow}
\usepackage{color}

\newcommand{\be}{\begin{equation}}  
\newcommand{\ee}{\end{equation}}  
\newcommand{\bea}{\begin{eqnarray}}  
\newcommand{\eea}{\end{eqnarray}}  

\DeclareRobustCommand{\fbi}{\ensuremath{\mathrm{fb}^{-1}}}

\newcommand{\specialcell}[2][c]{%
  \begin{tabular}[#1]{@{}c@{}}#2\end{tabular}}

\begin{document}

\vspace*{1.2cm}

\thispagestyle{empty}
\begin{center}
{\LARGE \bf Searching for axion-like particles with proton tagging at the LHC}

\par\vspace*{20mm}\par

{\large \bf C. Baldenegro$^a$, S. Fichet$^b$, G. von Gersdorff$^c$, C. Royon$^a$}

\bigskip

{\em $^a$ University  of Kansas, Lawrence, Kansas, U.S.}
\\
{\em $^b$ ICTP-SAIFR \& IFT-UNESP, R. Dr. Bento Teobaldo Ferraz 271, S\~ao Paulo, Brazil}
\\
{\em $^c$ Departamento de F\'isica, Pontif\'icia Universidade
Cat\'olica de Rio de Janeiro, Rio de Janeiro, Brazil }
\vspace*{5mm}

{\em E-mail: 
c.baldenegro@cern.ch\\
sylvain.fichet@lpsc.in2p3.fr\\
gersdorff@gmail.com\\
christophe.royon@cern.ch}

\vspace*{15mm}

{  \bf  Abstract }

\end{center}
\vspace*{1mm}

\noindent
\begin{abstract}

The existence of an axion-like particle (ALP) would induce anomalous  scattering of light by light. This process can be probed at the Large Hadron Collider in central exclusive production of photon pairs in proton-proton collisions by tagging the surviving protons using forward proton detectors. Using a detailed simulation, we estimate the expected bounds on the ALP--photon coupling for a wide range of masses. We show that the proposed search is competitive and complementary to other collider bounds for masses above 600 GeV, especially for resonant ALP production between 600 GeV and 2 TeV.  
 Our results are also valid for a CP-even scalar, and the efficiency of the search is independent of the width of the ALP.
\end{abstract}

\noindent
\clearpage

\section{Introduction}\label{sec:intro}

The presence of light (pseudo) scalars coupled to particles of the Standard Model (SM) of particle physics would have numerous consequences from the subatomic to the cosmological scale. These particles might address the longstanding question of why quantum chromodynamics seems to not break the CP symmetry \cite{Peccei:1977hh,PhysRevLett.38.1440}, as well as explain a possible component of dark matter. Axion-like particles (ALPs) appear in many extensions of the SM. For example, CP-odd scalars typically appear in the string theory landscape \cite{Witten:1984dg,Conlon:2006tq,Svrcek:2006yi,Arvanitaki:2009fg, Acharya:2010zx, Cicoli:2012sz}, 
 or theories with spontaneously broken approximate symmetries \cite{Masso:1995tw}. A CP-even scalar can for instance be the radion mode from an extra dimension~\cite{Csaki:2007ns}, the dilaton arising from spontaneous breaking of conformal symmetry~\cite{Goldberger:2008zz}, or the radial mode of the symmetry-breaking vacuum in composite Higgs models~\cite{Fichet:2016xvs}. Any new neutral spin-0 particle added to the SM typically couples to fermions only via dimension-five operators proportional to the fermion mass, while its dominant coupling to gauge bosons is via dimension-five operators containing derivatives. Therefore, at energies above the top quark mass, these particles are mostly accessible via their couplings to gauge bosons or the Higgs boson. In this paper, we are primarily interested in their coupling to photons.

These particles have been strongly constrained by numerous observations\footnote{An up-to-date review can be found in Ref. \cite{Bauer:2017ris}.}, some of them by dedicated experiments. A number of these constraints are model-independent, in the sense that they can be shown in the plane of mass versus (pseudo)scalar--photon coupling.
This landscape of constraints tends to vanish at high masses, where searches are collider-based. Indeed, searching for a particle relying only on its coupling to photons is not an easy task at a lepton or hadron collider.
Bounds obtained from the LEP, Tevatron and LHC in tri-photon, di-photon and mono-photon final states reach the highest masses, but are not very sensitive. Conversely, bounds from on-shell Higgs and $Z^0$ boson decays into an ALP in association with visible SM particles have been studied recently in Ref. \cite{Bauer:2017ris} along with their projections for Run-2 of the LHC. The sensitivity found is very good, but the mass reach is necessarily bounded from above by kinematics.

A particularly interesting proposal to study the ALP--photon coupling is via elastic scattering of light-by-light in ultraperipheral heavy-ion collisions, as pointed out in Ref.~\cite{Knapen:2016moh}. In these collisions, the coherent photon-photon luminosity is proportional to $Z^4$ ($Z=82$ for lead), which enhances significantly the cross section for exclusive diphoton production. Evidence for light-by-light scattering in ultraperipheral heavy-ion collisions was reported by the ATLAS collaboration~\cite{Aaboud:2017bwk} and the corresponding bounds on the ALP--photon coupling were derived shortly afterwards~\cite{Knapen:2017ebd}.
These bounds can be quite robust for masses from 1 GeV to 100 GeV, with its reach in mass limited from above by the minimum impact parameter in ultraperipheral heavy-ion collisions. However, for larger masses usually accessible in proton-proton (p-p) collisions at the LHC, the ALP search remains very challenging.

\begin{figure}[t]
\centering
\includegraphics[width=0.4\textwidth]{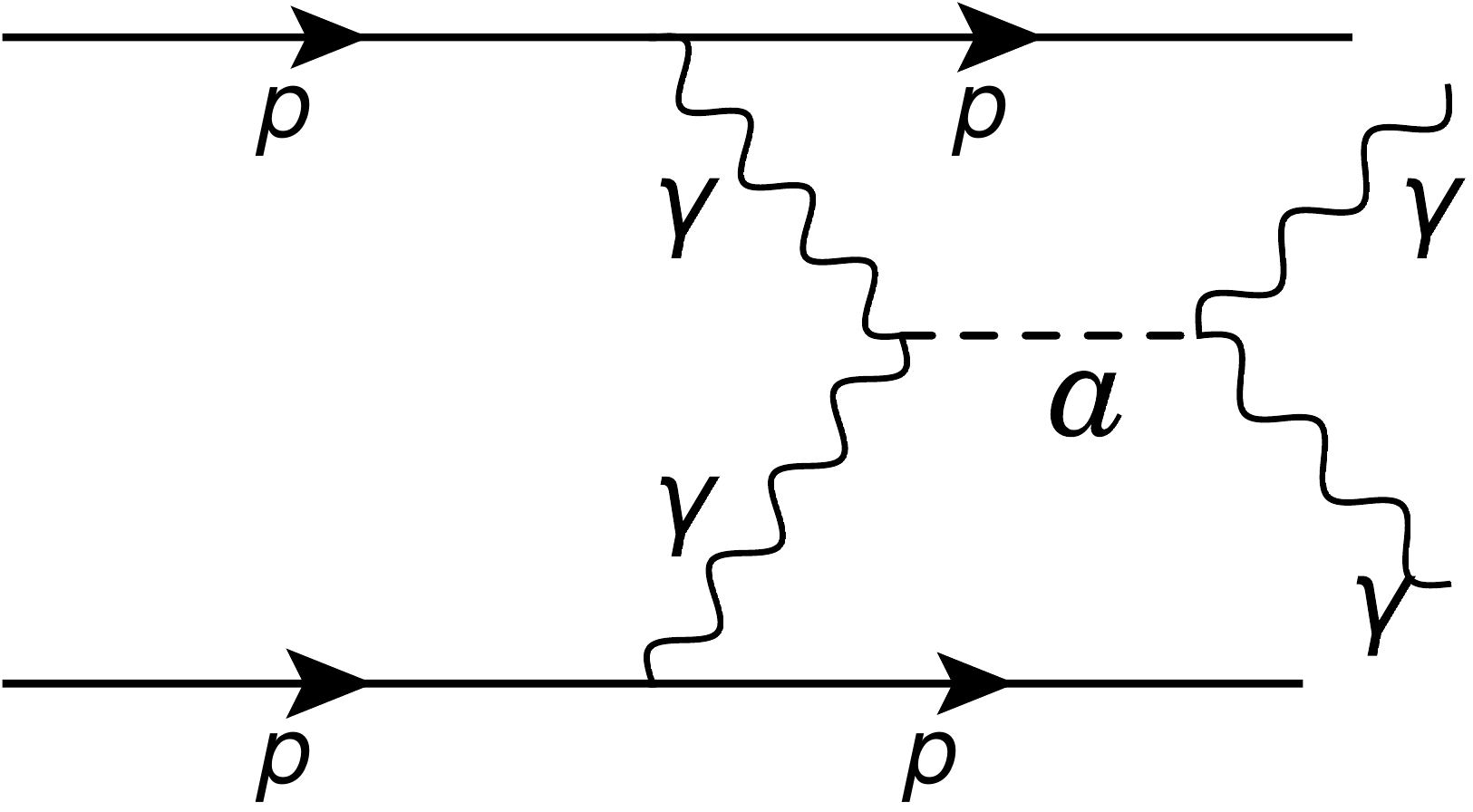}
\caption{\label{exclusive_diphoton} Schematic diagram of an axion-like particle production in two-photon coherent emission in proton-proton collisions. The scattered intact protons are tagged with the forward proton detectors and the photon pair is detected in the central detector.}
\end{figure}

\begin{figure}[b]
\centering
\includegraphics[width=0.9\textwidth]{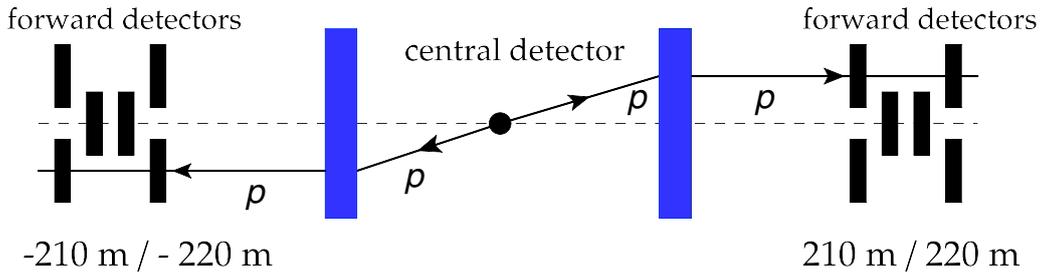}
\caption{\label{proton_tagging} Schematic diagram of the proton tagging method at the LHC in central exclusive processes. The central detector (circle) collects the photon pair. The LHC magnets (blue) act as a precise momentum spectrometer on the outgoing intact protons. The protons pass through the forward detectors (black boxes) and their kinematic information is reconstructed offline. The dashed line represents the beamline.}
\end{figure}

This paper presents an extension of the search for a (pseudo) scalar in light-by-light scattering at the LHC. 
 We propose to search for ALPs in central exclusive diphoton production in p-p collisions (see Fig. \ref{exclusive_diphoton}),
\be
pp\rightarrow p(\gamma \gamma \rightarrow \gamma\gamma) p
\ee
where the photon pair is measured in the central detector and the scattered intact protons are tagged with dedicated forward proton detectors, which are installed symmetrically at a distance of about 210 m (220 m) with respect to the interaction points of the CMS (ATLAS) experiment (see Fig. \ref{proton_tagging} ). Using proton tagging, we can reach diphoton invariant masses between 350 GeV and 2 TeV, where the acceptance of the forward detectors is nearly 100\% efficient.

 The LHC magnets around the interaction points of CMS and ATLAS act as a precise longitudinal momentum spectrometer on the protons that have lost a fraction of their original momentum due to the photon exchange. The forward proton detectors are equipped with charged particle trackers to tag the intact protons. The proton fractional momentum loss $\xi = \Delta p /p$ is reconstructed offline.
 Compared to other exclusive production searches, which usually rely on vetoes on the detector activity (for example, absence of calorimeter activity in the forward and backward rapidities above a threshold), the proton tagging method directly measures the proton surviving the coherent photon emission.
 
  The forward detectors, together with the central detector, enable the complete reconstruction of the collision event. This sets a kinematical constraint on the final state that allows an efficient offline selection for central exclusive processes with large background rejection factors.
Other studies based on proton tagging at the LHC 
for new physics searches can be found in 
\cite{usww, usw,Sahin:2009gq,Atag:2010bh, Gupta:2011be, Epele:2012jn, Lebiedowicz:2013fta, Fichet:2013ola,Fichet:2013gsa,Sun:2014qoa,
Sun:2014qba,Sun:2014ppa,Sahin:2014dua,Inan:2014mua,
Fichet:2014uka,Fichet:2015nia,Cho:2015dha,Fichet:2016clq,Fichet:2015vvy,Fichet:2016pvq,
Baldenegro:2017aen,
Baldenegro:2017lzv}.

 The detectors capable of studying central exclusive production in Run-2 of the LHC are the ATLAS Forward Physics (AFP) \cite{Adamczyk:2017378} and the CMS-TOTEM Precision Proton Spectrometer (CT-PPS) \cite{Albrow:1753795}, which were brought online in 2017 and 2016 respectively. CT-PPS reported the first physics result at the LHC using the proton tagging method by observing the $pp\rightarrow p(\gamma\gamma\rightarrow \ell^+ \ell^-)p$ reaction \cite{Cms:2018het}, which serves as a proof of principle of the proton tagging method to study $\gamma\gamma$ collisions above the electroweak scale. About 40 fb$^{-1}$ of data has been recorded by AFP and CT-PPS for offline analysis at the time of writing this paper.

Qualitatively, the interest of our proposed search method compared to the ones mentioned at the beginning is that the mass reach is limited only by the center-of-mass energy of the proton-proton system and the acceptance of the forward detectors. The photon-photon luminosity for the $\gamma\gamma\rightarrow \gamma\gamma$ subprocess is much smaller than the one in an ultraperipheral heavy-ion collision, but on the other hand the reach in the center-of-mass energy of the photon pair is higher, which enhances the sensitivity for the ALP search since the (pseudo) scalar--photon coupling vertex comes from an effective dimension-5 operator and thus grows rapidly with the diphoton collision energy. 
Another advantage of the exclusive channel in proton-proton collisions is that it does not need to rely on a dedicated bump search in the diphoton spectrum. Therefore, the results we can obtain are valid for a very broad resonance, which could be missed by an analysis relying on the diphoton mass spectrum lineshape.

The paper is organized as follows. The ALP--induced light-by-light scattering rate is calculated in Sec.~\ref{sec:lightbylight}. Our analysis strategy is then detailed in Sec. \ref{sec:analysis}, and Sec.~\ref{sec:simulation} summarizes the simulation
results.  The statistical framework used for the projections is laid down in Sec.~\ref{sec:statistical}. Section~\ref{sec:results} contains our final results, where we show our projections in the ALP parameter space.

\section{The $pp\rightarrow p(\gamma\gamma\rightarrow \gamma \gamma)p$ process}\label{sec:lightbylight}

We compute the production rates for light-by-light scattering in proton-proton collisions using the equivalent photon approximation \cite{Terazawa:1973tb}. In this approximation, the electromagnetic field generated by the fast moving protons can be considered as an intense photon beam. The photons exchanged by the colliding protons are almost on their mass shell (low virtualities $Q^2$).

The hadronic cross section can be calculated as a convolution of the effective photon fluxes and the $\gamma\gamma\rightarrow\gamma\gamma$ subprocess matrix elements,


\be
\frac{\mathrm{d}\sigma}{\mathrm{d}\Omega}^{pp\rightarrow p(\gamma\gamma\rightarrow \gamma \gamma)p} = \int \frac{\mathrm{d}\mathcal{L}}{\mathrm{d} \hat{s}}^{\gamma\gamma} \, \frac{\mathrm{d}\hat{\sigma} }{\mathrm{d}\Omega}^{\gamma\gamma\rightarrow\gamma\gamma}\, \mathrm{d}\hat{s}\,,
\ee
where $\frac{\mathrm{d}\mathcal{L}}{\mathrm{d} \hat{s}}^{\gamma\gamma}$ is the two-photon effective luminosity spectrum, obtained after integrating the photon fluxes over their virtualities \footnote{We use the lowest virtuality allowed by kinematics $Q^2_{\mathrm{min}} = \frac{2 m_p^2 E_{\gamma}^2}{\sqrt{s} (\sqrt{s}/2-E_{\gamma})}$, where $\sqrt{s}$ is the center-of-mass energy of the p-p system. We take the maximum virtuality of $Q^2_\mathrm{max}=4$ GeV$^{2}$.} and energies at a fixed two-photon center-of-mass energy $\hat{s}$, and $\frac{\mathrm{d}\hat{\sigma} }{\mathrm{d}\Omega}^{\gamma\gamma\rightarrow\gamma\gamma}$ is the subprocess differential cross section. We use the photon flux computed from the proton elastic electromagnetic form factor \cite{Budnev}.

In our treatment we consider a survival probability $\langle \mathcal{S}^2 \rangle = 0.9$, which quantifies the probability that the protons remain intact after the coherent photon emission. Phenomeno\-logy studies suggest that $\langle \mathcal{S}^2 \rangle$ can have weak dependence on the invariant diphoton mass $m_{\gamma\gamma}$ and $p^{\gamma}_\mathrm{T}$ due to soft rescattering effects between the surviving protons~\cite{SuperChic2,Khoze:2017sdd}, and it might have values down to $\langle \mathcal{S}^2 \rangle = 0.6$ for masses close to 2 TeV for some photo-produced processes, where the acceptance of the forward detectors vanishes. Ultimately, this non-perturbative quantity has to be measured experimentally by the AFP and CT-PPS experiments. We should mention that in our study we do not take into account polarization effects of the initial-state photons, as is often done in the equivalent photon approximation. Initial-state photons polarizations could yield differences in the production cross section, since they might affect the aforementioned soft rescattering effects~\cite{Khoze:2002dc,SuperChic2,Khoze:2017sdd}. To the best of our knowledge, there are no explicit calculations for polarization effects in $pp\rightarrow p(\gamma\gamma\rightarrow\gamma\gamma)p$ scattering, although results for specific processes are known~\cite{Khoze:2002dc,SuperChic2}. For purposes of the expected bounds calculation, which will depend on the total yield of the signal rather than on differential features, we will neglect these effects. However, if an ALP is discovered in this channel, a proper treatment of initial-state photons polarizations would be useful to determine the nature of the new particle.

In order to describe the interaction of the (pseudo) scalar $a$ with photons we use the effective interaction models	
\be
{\cal L}^+=\frac{1}{f} a F_{\mu\nu}F^{\mu\nu} \quad \textrm{(CP-even)}\,, \quad
{\cal L}^-=\frac{1}{f} a F_{\mu\nu} \tilde F^{\mu\nu} \quad \textrm{(CP-odd)} \,,
\label{eq:coupling}
\ee
where $f^{-1}$ is the ALP--photon coupling and $\tilde F^{\mu\nu}=\frac{1}{2}\epsilon^{\mu\nu \rho \sigma}F_{\rho \sigma}$. The contributions to the $\gamma\gamma\rightarrow\gamma\gamma$ helicity amplitudes in both cases read
\be
{\cal M}_{++++}= - \frac{4}{f^2} \frac{s^2 }{s-m_a^2} \,,
\ee
\be
{\cal M}_{++--}= - (CP) \frac{4}{f^2} \left(\frac{s^2 }{s-m_a^2}+\frac{t^2 }{t-m_a^2}+\frac{u^2 }{u-m_a^2}\right) \,,
\ee
\be
{\cal M}_{+++-}= 0 \,,
\ee
and ${\cal M}_{+-+-}(s,t,u)={\cal M}_{++++}(u,t,s)$, ${\cal M}_{+--+}(s,t,u)={\cal M}_{++++}(t,s,u)$, where $s$, $t$ and $u$ are the Mandelstam variables of the diphoton system and $m_a$ is the mass of the ALP. This is an effective theory valid roughly up to energy $\sqrt{s}\sim 4\pi f$. The amplitude grows with $s$ and unitarity would be violated above this scale. The projections we derive in Sec. \ref{sec:results} yield coupling values well below this unitarity bounds. 
The unpolarized differential cross-section is given by 
\be
\frac{d \sigma}{d\Omega}=\frac{1}{128 \pi^2s} \left(|{\cal M}_{++++}|^2+
|{\cal M}_{+-+-}|^2
+|{\cal M}_{+--+}|^2
+|{\cal M}_{++--}|^2
 \right)\,.
 \label{eq:dsigma}
\ee
Each term in Eq.~(\ref{eq:dsigma}) in principle contains the SM light-by-light background. However, the production rate for this process is very small within the acceptance of the forward detectors. In particular, the CP odd and even cases yield the same differential cross section. 
The scalar being coupled to photons, it has a minimal decay width of
\be
\Gamma (a\rightarrow \gamma\gamma) = \frac{m_a^3}{4\pi f^2}\,.
\label{eq:pwidth}
\ee
In our upcoming projections the decay width of $a$ is a free parameter satisfying $\Gamma\geq\Gamma (a\rightarrow \gamma\gamma)$. The decay width will be parametrized via the branching ratio into photons ${\cal B}(a\rightarrow \gamma\gamma)=\Gamma (a\rightarrow \gamma\gamma)/\Gamma$.

It is instructive to examine the amplitudes when taking into account our knowledge of the forward detectors. The forward detectors have access to the process in a given interval of center-of-mass energy $\sqrt{s}\in[\sqrt{s}_0,\sqrt{s}_1]$, where $\sqrt{s}_0$ is sizeable. For the forward detectors installed at ATLAS and CMS we have roughly $\sqrt{s}_0\sim 350$ GeV, $\sqrt{s}_1\sim 2$ TeV, where the acceptance is efficient. It follows that one can distinguish three regimes for the diphoton production rate. 

If $m_a<\sqrt{s}_0$, the mass of the particle is negligible from the viewpoint of the detectors, hence the sensitivity will be independent of $m$ and of the width of the particle. 
 Note that the cross section still grows as $s$ in this regime as a consequence of  the
  non-renormalizability of the interaction Eq.~(\ref{eq:coupling}), hence the search for a light particle coupled to photons can  benefit a lot from an increase in collision energy.  
 If $\sqrt{s}_0<m_a<\sqrt{s}_1$, the scalar is produced resonantly. In that regime a bump search can be performed, unless the resonance is very broad. When the resonance is narrow, the cross section behaves as \be\sigma_{\gamma\gamma\rightarrow a\to\gamma\gamma} \propto f^{-2} {\cal B}_{\, a \rightarrow \gamma\gamma }\,.\ee
 Finally if $m_a>\sqrt{s}_1$, the scalar is too heavy to be produced resonantly within the acceptance of the detector. Taking the $m_a\gg \sqrt{s}$ limit, the amplitude can then be described by the low-energy effective field theory 
%
\be
\mathcal L^+_{\rm eff}=\frac{1}{2f^2m_a^2}(F_{\mu\nu}F^{\mu\nu})^2\,,\qquad
\mathcal L^-_{\rm eff}=\frac{1}{2f^2m_a^2}(F_{\mu\nu}\tilde F^{\mu\nu})^2\,,\qquad
\ee
These three regimes will clearly appear on the sensitivity plots.

\section{Analysis framework}\label{sec:analysis}

Our event selection treatment follows the method used in Ref.~\cite{usww, usw,Fichet:2014uka,Fichet:2015nia,Fichet:2016clq,
Baldenegro:2017aen,
Baldenegro:2017lzv} and resembles analyses reported by the ATLAS and CMS collaborations on search for exclusive diphoton production in p-p and Pb-Pb collisions \cite{Aaboud:2017bwk,Chatrchyan:2012tv}. We consider proton-proton collisions at a center-of-mass energy of 13 TeV and an integrated luminosity of 300 fb$^{-1}$. We look for photons reconstructed in the barrel region $|\eta|<2.5$, where the reconstruction efficiency is on average 80\% for energetic photons \cite{Khachatryan:2015iwa,MITREVSKI20162539}. The photon energy resolution $\Delta E^\gamma / E^\gamma$ is taken as 1\% since we are dealing with multi-GeV photons. We ask for the leading (subleading) photon to have a minimum transverse momentum of 200 (100) GeV. To better isolate elastically produced photon pairs, we apply a cut on the azimuthal angle separation between the two photons $|\Delta\phi^{\gamma\gamma}-\pi| < 0.01$ and their transverse momentum ratio $p^{\gamma}_\mathrm{T,2}/p^{\gamma}_\mathrm{T,1} > 0.95$. We verified the stability of the elastic selection on the signal by varying the azimuthal angle separation between the two photons as $|\Delta\phi^{\gamma\gamma}-\pi| < 0.04$ and $p^{\gamma}_\mathrm{T,2}/p^{\gamma}_\mathrm{T,1}>0.90$. This diphoton selection yields an acceptance of about 80\% on the signal. Finally, we apply a cut on the invariant mass of the photon pair of 600 GeV for background suppression purposes. We assume that the trigger efficiency is close to 100\% at the end of our offline selection.

Since the forward detectors can not get arbitrarily close to the proton beam, and the position of the LHC beam collimators limits the acceptance of the forward detectors from above, the resulting design acceptance on the protons fractional momentum loss is $0.015 \leq \xi \leq 0.15$. We assume that $\xi$ is known to 5\% precision.

The backgrounds for exclusive photon pair production in p-p collisions can be classified in reducible and irreducible backgrounds. The irreducible background comes from the SM light-by-light scattering process, which is induced at one-loop at leading order. This background is greatly reduced within the mass acceptance of the forward proton detectors. Two-gluon exchange between the two colliding protons can lead to a photon pair with intact protons in the final state. However, this background is suppressed more rapidly compared to the SM light-by-light scattering at larger invariant diphoton masses are \cite{Fichet:2014uka}, since we are asking for no additional gluon real emission from the gluon ladder diagram, which formally introduces a Sudakov suppresion factor \cite{Khoze:2001xm}. Finally, we consider central exclusive prodution of $e^+e^-$, where the dielectron is misidentified as a photon pair. The misidentification rate electron is about 1\%. We find that in our final selection the expected central exclusive production is negligible.

\begin{figure}
\centering
\includegraphics[width=.8\textwidth]{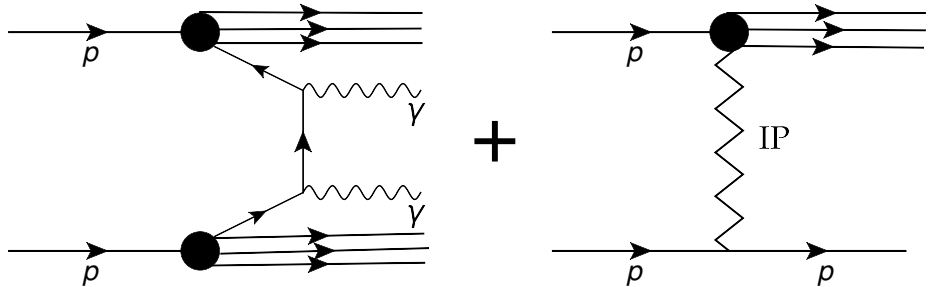}
\caption{\label{diagram_reducible} The dominant background for central exclusive diphoton production comes from non-exclusive photon pair production (left) overlapped with uncorrelated protons coming from soft diffractive processes in the additional interactions per bunch crossing (right).}
\end{figure}

The dominating background is the overlap of a non-exclusive photon pair and uncorrelated protons coming from soft diffractive interactions (see Fig. \ref{diagram_reducible}). Protons originating from soft diffractive processes can have fractional momentum losses $\xi$ populating the signal region, and can lead to fake signals. The cross section for diffractive interactions is known to be very large (about 1 mb), and the number of secondary interactions per bunch crossing (pileup) at the current instantaneous luminosity at the LHC in p-p collisions (50 interactions per bunch crossing at worst) enhances the likelihood of faking the signal.
However, central exclusive production events satisfy $m_{\gamma\gamma} = \sqrt{\xi_1\xi_2 s}$ and $y_{\gamma\gamma} = \frac{1}{2}\log(\frac{\xi_1}{\xi_2})$. Thus, we apply a cut $|\sqrt{\xi_1\xi_2 s}/m_{\gamma\gamma}-1|<0.03$ and $|y_{\gamma\gamma} - \frac{1}{2}\log(\frac{\xi_1}{\xi_2})|<0.03$. This last point is the key to the precision of this search, as it suppresses a large portion of the background, since non-exclusive photon pairs and protons arising from soft diffractive processes are kinematically uncorrelated.

Other processes contributing to the reducible background are misidentified photon pairs overlapped with soft diffractive protons. We consider non-exclusive $e^+e^-$ pair production (Drell-Yan), since electrons and positrons can fake the diphoton pair, and non-exclusive dijet production, since hard partons can hadronize into a large number of $\pi^0$ mesons which subsequently decay into photon pairs, which can fake the photon detection.

Finally, there is a semi-exclusive contribution from diphoton production in double pomeron exchange. This color-singlet exchange can lead to surviving protons and two energetic photons in the final state. However, the production yield falls rapidly as a function of the diphoton invariant mass, and the intact protons and the diphoton are not strongly correlated kinematically since part of the energy is carried by the hadronization of the pomeron. This contribution is negligible in our final selection.

After applying the offline event selection described in this section, we end up with an almost background-free probe for light-by-light scattering in p-p collisions at high diphoton invariant masses, sensitive to cross sections as small as a fraction of a fb, as found before in Refs. \cite{Fichet:2014uka,Baldenegro:2017aen}.

\section{Simulation results}\label{sec:simulation}

The signal $\gamma\gamma\rightarrow a \rightarrow \gamma\gamma$ subprocess was implemented and generated with the Forward Physics Monte Carlo event generator (FPMC) \cite{FPMC}. FPMC is an event generator for diffractive and photon-induced processes in hadronic collisions. The SM light-by-light scattering process is also simulated in FPMC, which includes contributions from charged leptons and the $W$ boson in the one-loop diagram. We also simulated exclusive dielectron production with this generator.

We also employed FPMC for the double pomeron exchange background. FPMC uses the fits based on H1 inclusive diffractive results \cite{Royon:2006by} of the pomeron in the Ingelman-Schlein \cite{INGELMAN1985256} pomeron parton distribution functions parametrization. The diffractive parton distribution functions are convoluted with the hard scattering processes library in HERWIG 6.5 \cite{herwig}. FPMC includes a survival probability of 0.03~\cite{survival1,survival2} for pomeron exchange processes. It is not known what the pomeron flux would be at 13 TeV, as it has to be constrained experimentally. One may consider this as a theoretical uncertainty. We recomputed the background yield by rescaling it by a factor of 10 and 100. In neither of these cases we see a significant contamination after the final selection.

Non-exclusive backgrounds, which include diphoton production, dijet production and $e^+e^-$ in Drell-Yan, are simulated in PYTHIA8 \cite{Sjostrand:2007gs} using the parton distribution function set NNPDF2.3 QCD+QED LO parametrization. We verified that our results do not depend on the choice of the parton distribution function set. For the misidentified jets, we use the anti-$k_\mathrm{T}$ algorithm with the FastJet package \cite{Fastjet} with a cone radius $R=0.4$.
The probability of tagging at least one proton per diffractive interaction is estimated from the minimum bias library of PYTHIA8 \cite{Timing}. We assume that the number of secondary interactions per bunch crossing at the interaction points of the LHC follow a Poisson distribution with mean $\mu = 50$, the largest number of interactions per bunch crossing at the LHC in Run-2. For each non-exclusive photon pair event, we sample the proton tag probability estimated before for protons coming from soft diffractive events from the additional interactions. If there is at least one proton tagged on each arm, we draw the corresponding $\xi$ value from a parametrization of the differential yield of diffractive protons $\frac{\mathrm{d}N}{\mathrm{d}\xi}^\mathrm{diff} \sim \frac{1}{\xi}$. If more than one proton is present in a given arm, we choose the one that yields the closest mass and rapidity to the diphoton system.

\begin{figure}
\centering
\includegraphics[width=.65\textwidth]{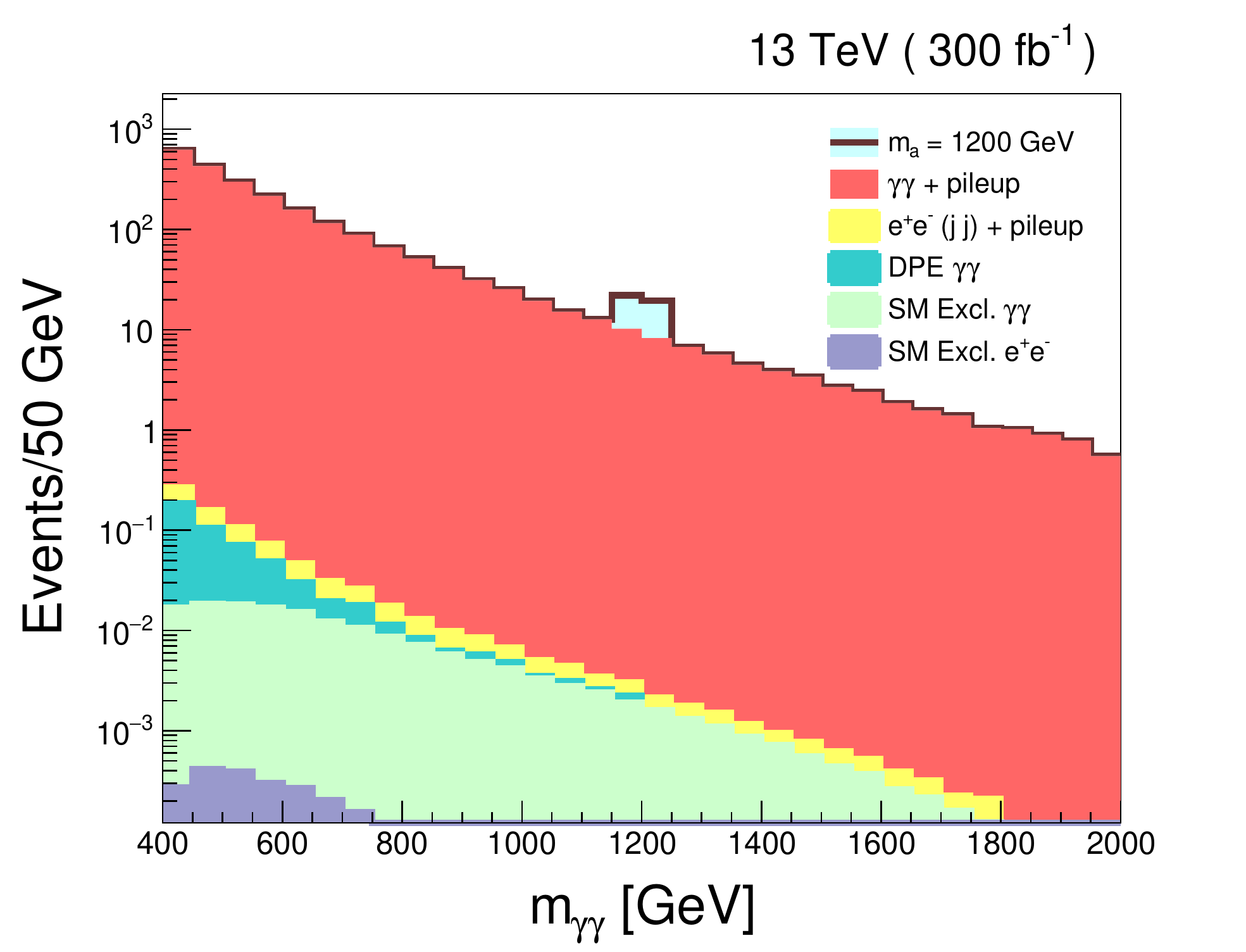}
\caption{\label{fig:diphoton_mass} Differential yield as a function of the photon pair invariant mass for exclusive diphoton candidates with two tagged protons within the acceptance $0.015<\xi_{1,2}<0.15$. No elastic or exclusive offline selection is applied for the diphoton candidates in this plot. We assume there are in average 50 secondary interactions per bunch crossing. For illustrative purposes, we show an instance of a resonant ALP production with $m_{a} = 1200$ GeV and a coupling value $f^{-1} = 0.1$ TeV$^{-1}$.}
\end{figure}

\begin{figure}
\centering
\includegraphics[scale=0.35]{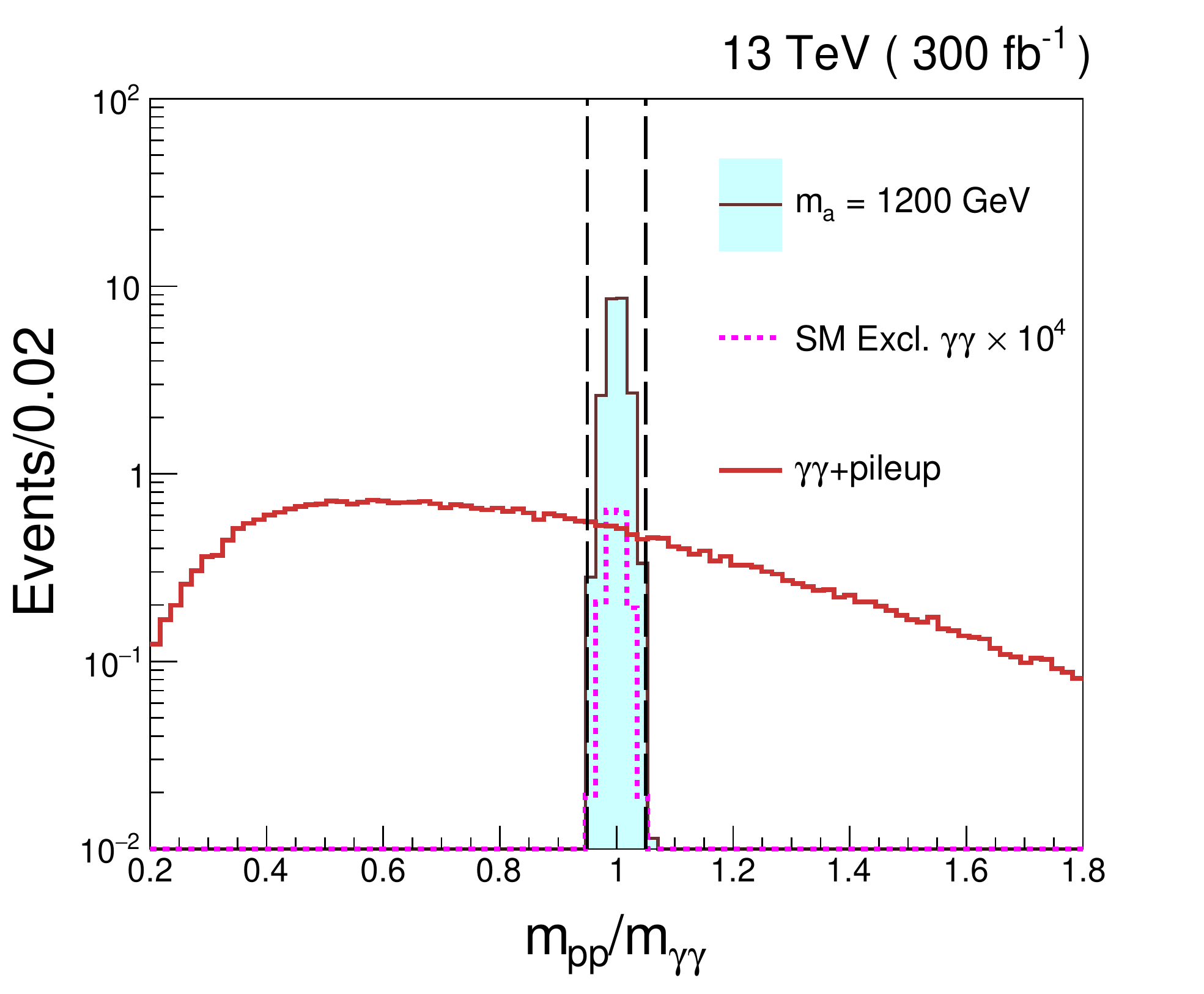}
\includegraphics[scale=0.35]{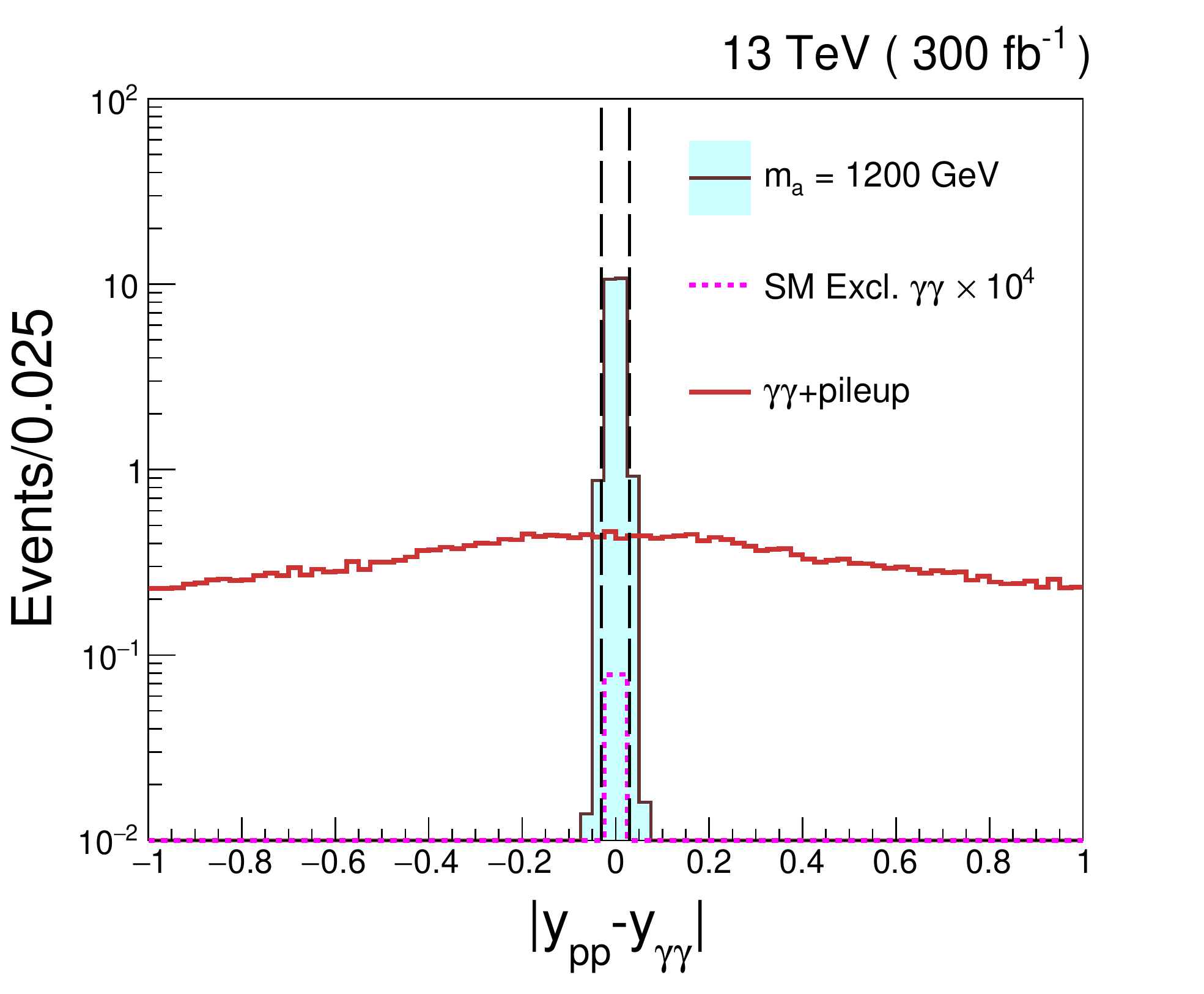}
\caption{\label{fig:exclusive_selection} Distributions of the ratio of the diphoton mass reconstructed with the forward detectors $m_{pp} = \sqrt{\xi_1\xi_2 s}$ to the reconstructed diphoton mass $m_{\gamma\gamma}$ (left) and the difference of the diphoton rapidity $y_{\gamma\gamma}$ and the rapidity reconstructed with the forward detectors $y_{pp}=\frac{1}{2}\log(\frac{\xi_1}{\xi_2})$ distribution (right). Diphoton candidates in these plots have passed the elastic selection and the mass lowerbound of 600 GeV.  A strong correlation between the forward-backward and central information can be seen for the signal (light blue), while for the background (red line) we see these variables are uncorrelated. We select diphoton candidates lying inside the dashed vertical lines. The width of the signal in these plots is caused mainly by the $\xi_{1,2}$ resolution. The integrated luminosity is 300 fb$^{-1}$ and the average number of pileup interactions is $\mu = 50$. The intact protons lie within the acceptance $0.015<\xi_{1,2}<0.15$.}
\end{figure}

The signal and background yields after the sequential selection cuts described in Sec. \ref{sec:analysis} can be seen in Tab. \ref{tab:cutflow}. After all the selection cuts, we end up with a near background-free probe of light-by-light scattering in p-p collisions. The differential yield for the exclusive diphoton candidates can be seen in Fig. \ref{fig:diphoton_mass}. The high signal selection efficiency of the exclusive selection is illustrated in Fig. \ref{fig:exclusive_selection}.

\begin{table}[t]
\centering
\small
\begin{tabular}{|c||c||c|c|c|c|}
\hline
Sequential selection & \specialcell{ALP} & Excl. SM & DPE $\gamma\gamma$ & \specialcell{$e^+e^-$ / dijet \\ +pileup}  & \specialcell{$\gamma\gamma$\\+ pile up} \\
\hline
\hline
\specialcell{$[0.015<\xi_{1,2}<0.15$, \\ $p_{\mathrm{T}1,(2)}>200,(100)$ GeV]}       & 23.1   & 0.1 & 0.1  & 1.2           & 1246         \\ \hline
$m_{\gamma\gamma}>600$~GeV                                                           & 23.1   & 0.06 &  0    & 0.1          & 440         \\ \hline
\specialcell{[$p_{\mathrm{T2}}/p_{\mathrm{T1}}>0.95$,\\ $|\Delta\phi^{\gamma\gamma}-\pi|<0.01$]}    & 23.1    & 0.06 & 0   & 0         & 35         \\ \hline
$|m_{pp}/m_{\gamma\gamma}-1| < 0.03$                                  & 21.8   & 0.06 & 0   & 0         & 1.2          \\ \hline
$|y_{\gamma\gamma}-y_{pp}|<0.03$                                                     & 21   & 0.06 & 0   & 0         & 0.2            \\
\hline
\end{tabular}
\caption{Signal and background yields after applying the event sequential selections. For illustrative purposes, we choose an ALP with mass $m_a = 1200$ GeV and a coupling value of $f^{-1} = 0.1$ TeV$^{-1}$. We assume an integrated luminosity of 300~\fbi\, an average of $50$ additional interactions per bunch crossing at $\sqrt{s}=13$ TeV. Excl. stands for the exclusive backgrounds and DPE for double pomeron exchange background. Non-exclusive diphoton overlapped with soft diffractive protons (rightmost column) constitute the dominating background. The first two rows correspond to the diphoton offline preselection. The third row corresponds to the elastic selection. The last two rows correspond to the exclusive selection, with $m_{pp} = \sqrt{\xi_{1}\xi_{2}s}$ and $y_{pp} = \frac{1}{2}\log(\frac{\xi_1}{\xi_2})$.
}
\label{tab:cutflow}
\end{table}

\clearpage

\clearpage

\section{Statistical framework}\label{sec:statistical}

For our projections, we need to assume a set of observed data. As commonly done, we assume that no statistical fluctuations are present in these imaginary data, which are usually dubbed ``Asimov'' data and that we denote here with a prime. As mentioned before, the integrated luminosity is chosen to be $L = 300$\,fb$^{-1}$, which is about the expected integrated luminosity to be taken by CT-PPS and AFP in Run-2.

The observed events follow a Poisson distribution and for this kind of analysis we can safely neglect the systematic uncertainties. 
The statistics for the events together with the prediction for the event rates  sets the likelihood function needed for our analysis,
\be
{\cal L}(\sigma )={\rm Pr}( n' |b+\sigma L)\,,\quad {\rm Pr}( \hat n |n)=\frac{ {n^{\hat n}} e^{-n  } }{\hat n !}\, \label{eq:L}
\ee
where $b$ is the expected number of events from the background. Thanks to the forward detectors, $b$ is very small, $b\approx 0.2$ for $L=300$\,fb$^{-1}$.  

Our method to obtain the projected $95\%$ CL exclusion contours is the following. We define the posterior probability density for $\sigma$ as $ {\cal L}(\sigma ) \pi(\sigma)$ where the prior  is $\pi(\sigma)=1$ if $\sigma>0$, and $0$ otherwise.  In order to obtain exclusion contours we will assume that no event is observed, \textit{i.e.} $n'=0$. This is a reasonable scenario because the expected background is $b\approx 0.2$.  
  The non-observation of events sets an upper bound on the signal event rate.  The higher posterior density region at $1-\alpha $ credibility level is solved analytically and is simply given by
\be
1-\alpha=\frac{\int^{\sigma_\alpha}_0 L(\sigma) \pi(\sigma)}
{\int_{0}^\infty L(\sigma)\pi(\sigma)}= 1-e^{-\sigma_\alpha L}\,,
\ee
hence the boundary of the exclusion region is given by  \be
\sigma_\alpha=-\frac{1}{L} \log(\alpha)\,.
\ee
Therefore for a $95\%$ credible interval, one takes $\alpha=0.05$ and the exclusion limit is simply given  by $\sigma_\alpha \approx 3 L^{-1}$.

Let us finally comment on the information from signal and background differential distributions. Having resonant production in some regime, it is in general useful to use the characteristic lineshape of a resonant signal to identify it over a background. When doing such ``bump search'', a likelihood comparing signal and background for every bin has to be used (see \textit{e.g.} \cite{Ferreira:2017ymn}). However, as long as we consider the case $n'=0$, there is no need to perform such shape analysis, which would require a precise evaluation of the background on each bin.  This is because when $n'=0$, the likelihood on each bin is an exponential, hence the likelihood for shape analysis becomes equivalent to the likelihood of  Eq.~\eqref{eq:L}.


\clearpage

\section{Results and discussion}\label{sec:results}

\begin{figure}
\centering
\includegraphics[width=1.\textwidth]{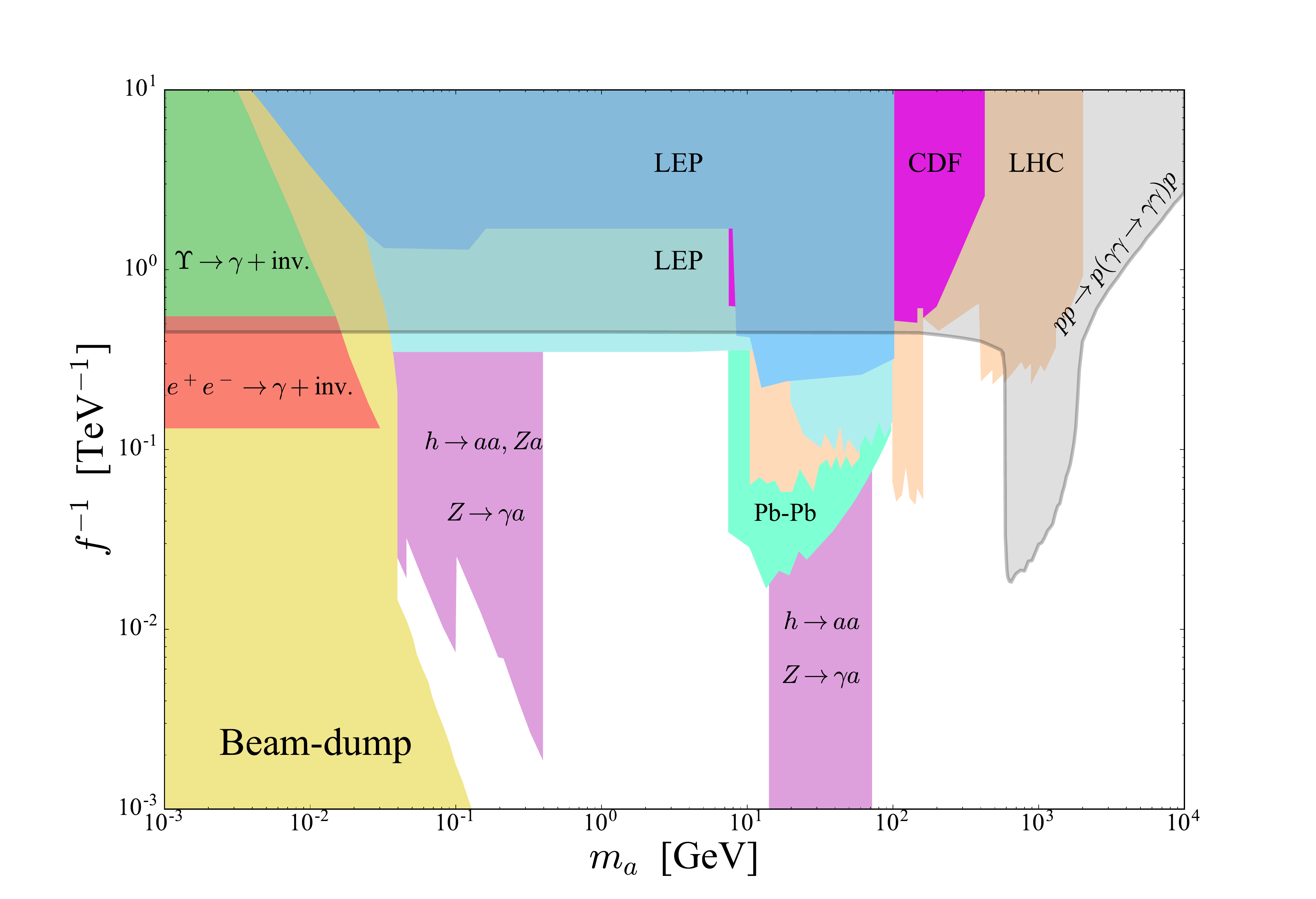}
\caption{\label{axion_reach} Exclusion regions on the ALP--photon coupling $f^{-1}$ and mass of the ALP $m_a$ plane. On light-shaded grey, we have the expected $95\%$CL exclusion limit in central exclusive diphoton production events assuming $\mathcal{B}(a\rightarrow \gamma\gamma) = 1$ for 300 fb$^{-1}$ in Run-2 of the LHC. Existing bounds are in solid color, are extracted from Ref.~\cite{Bauer:2017ris}, and can depend on additional assumptions (see text for details).}
\end{figure}

\begin{figure}[t]
\centering
\includegraphics[width=0.8\textwidth]{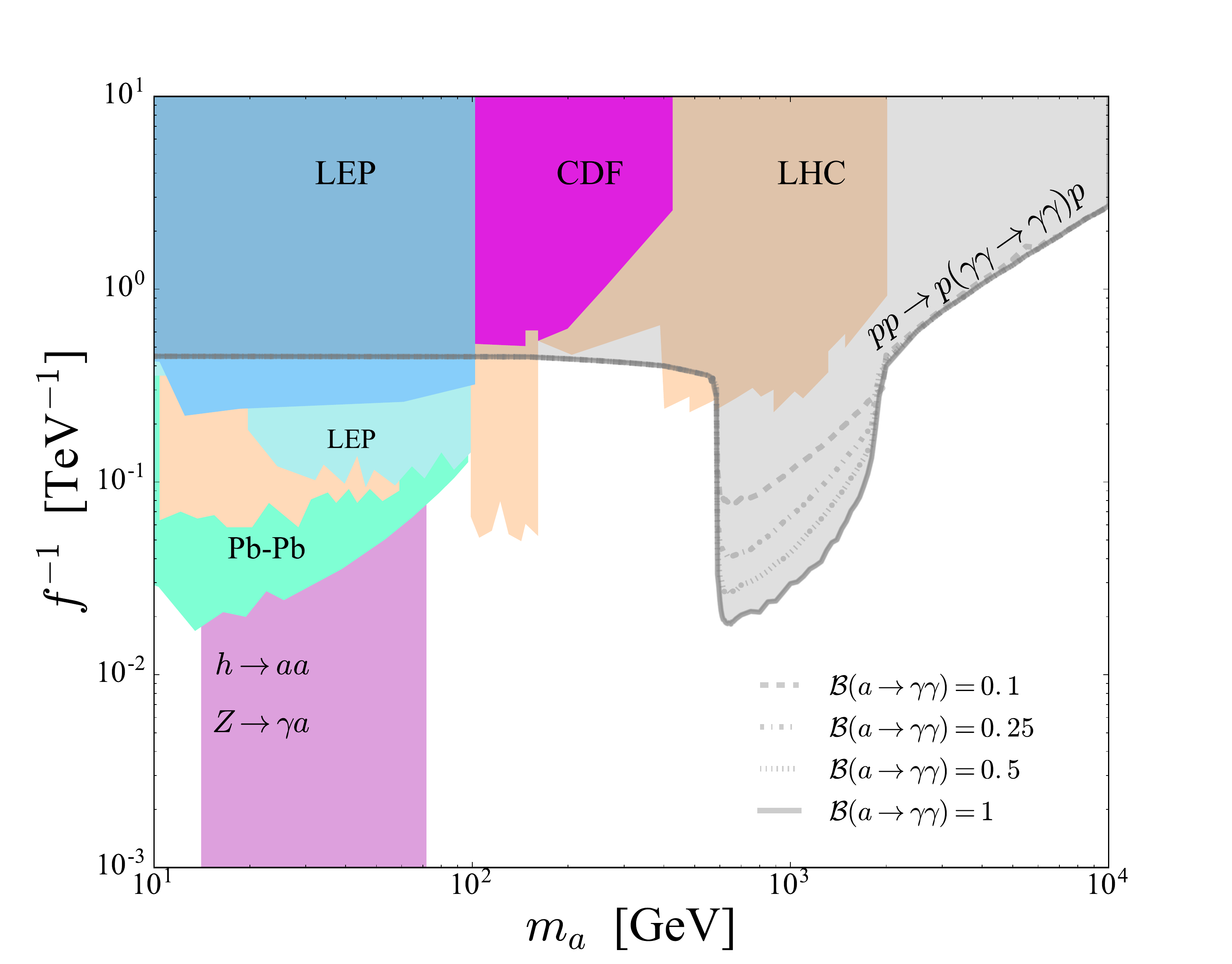}
\caption{\label{axion_reach_br} Exclusion regions on the ALP--photon coupling $f^{-1}$ and mass $m_a$ plane. On light-shaded grey, we have the expected $95\%$CL exclusion limit for 300 fb$^{-1}$ in central exclusive diphoton production events for different branching ratios of the ALP into two photons. Current excluded regions for $\mathcal{B}(a\rightarrow\gamma\gamma)=1$ are in solid color, are extracted from Ref.~\cite{Bauer:2017ris}, and can depend on additional assumptions (see text for details).}
\end{figure}

The expected sensitivity from the exclusive diphoton search can be represented in the $m_a-f$ plane. No other assumption is needed except in the region of resonant production where the branching ratio into photons has to be fixed.
The expected bound is displayed in Fig. \ref{axion_reach} in the ALP--photon coupling and mass plane for a centrally produced ALP with branching ratio $\mathcal{B}(a\rightarrow\gamma\gamma) = 1$. The lowest coupling values range between 0.02 TeV$^{-1}$ and 0.06 TeV$^{-1}$ for masses between 600 GeV to 1.5 TeV. The bound increases rapidly from 1.5 TeV to 2 TeV and follows a power-law-like behavior for masses larger than 2 TeV independently of the particle width. For masses below 600 GeV, the coupling is independent of the particle width and has a value of about 0.4 TeV$^{-1}$. Bounds with different fixed branching ratios can be seen in Fig. \ref{axion_reach_br}.

We extracted the existing exclusion limits \footnote{The ALP--photon coupling in \cite{Bauer:2017ris} is related to our coupling convention via $f^{-1} = \frac{C^{\mathrm{eff}}_{\gamma\gamma}}{\Lambda} e^2$.} from the  compilation in Ref.~\cite{Bauer:2017ris}. We show a subset of these bounds for masses between 10$^{-3}$ GeV and 2 TeV and coupling values as low as $10^{-3}$ TeV$^{-1}$, although this landscape  of bounds exists down to values of 10$^{-15}$ GeV in mass and 10$^{-10}$ TeV$^{-1}$ in the coupling. Beam dump searches probe resonant production of neutral pseudoscalar mesons in photon interactions with nuclei (Primakoff effect). Different beam dump runs at SLAC collectively yield the area in yellow \cite{PhysRevLett.59.755,PhysRevD.38.3375,Döbrich2016}. Upsilon meson decays searched at the CLEO and BaBar experiments \cite{PhysRevD.51.2053,delAmoSanchez:2010ac} exclude the region shaded in  green.
Bounds from collider-based searches for ALPs include measurements of mono-photons with missing transverse energy  ($e^+e^- \rightarrow \gamma$+invisible) at the LEP (orange), tri-photon searches on and off the $Z^0$ pole ($e^+e^-\rightarrow 3\gamma$) at the LEP (light blue and dark blue), and searches for the same final states in p-$\bar{\mathrm{p}}$ collisions at CDF (magenta) and in p-p collisions at the LHC (peach). The derivation of these collider-based bounds are discussed in detail in Ref. \cite{Jaeckel:2012yz,Mimasu:2014nea,Jaeckel:2015jla,Knapen:2016moh}. The region labelled as ``Pb-Pb'' (light green) was derived in Ref. \cite{Knapen:2017ebd} based on the measurement of light-by-light scattering in ultraperipheral heavy-ion collisions by the ATLAS collaboration \cite{Aaboud:2017bwk}. These collider-based bounds assume $\mathcal{B}(a\rightarrow\gamma\gamma)=1$. We also include recent constraints presented in Ref. \cite{Bauer:2017ris} based on Higgs boson and $Z$ boson exotic decays $h\rightarrow Za$, $h\rightarrow aa$ and $Z\rightarrow \gamma a$, where $a$ decays into a pair of charged leptons or a photon pair, which collectively exclude the parameter space in lavender. These bounds are based on measurements at the LHC in $h\rightarrow Z\gamma$, $h\rightarrow \gamma\gamma$, $h\rightarrow 4\gamma$ and $Z\rightarrow 3\gamma$ final states. Unlike our bounds, these constraints assume a given coupling value of the ALP to the Higgs boson and $Z$ boson.

The LHC region at high mass comes from CMS and ATLAS bump searches in the $\gamma\gamma$ spectrum recasted for ALPs in Ref.~\cite{Jaeckel:2012yz}. A basic extrapolation at 300 fb$^{-1}$ and $\sqrt{s}=13$~TeV luminosity tells that this LHC  exclusion region would be improved by a factor $\sim 4-5$. In the $0.6-2$\,TeV mass region, this is still below the expected sensitivity of our exclusive diphoton search by a factor $\sim 3-4$, showing that our method is competitive with respect to standard bump searches.

For $\mathcal{B}(a\rightarrow\gamma\gamma) < 1$, the sensitivity of our exclusive diphoton search in the $0.6-2$\,TeV mass range decreases as shown in Fig.~\ref{axion_reach_br}. This decrease comes simply from the lower signal rate. The total decay width increases for decreasing $\mathcal{B}(a\rightarrow\gamma\gamma)$, (as $\Gamma(a\rightarrow\gamma\gamma)$ is fixed by Eq.~(\ref{eq:pwidth})), but the efficiency of our search is width-independent as it consists of a mere counting of the total event number.  The region from LHC bump searches shrinks similarly when lowering $\mathcal{B}(a\rightarrow\gamma\gamma)$.  However, these searches are valid only for a narrow enough resonance, typically $\Gamma \lesssim 0.05\, m_a$ \cite{Jaeckel:2012yz}. 
  For sufficiently small $\mathcal{B}(a\rightarrow\gamma\gamma)$, the resonance width exceeds  this threshold value in the LHC exclusion region at large $m_a$, causing the bump search to lose its power. As a consequence our exclusive diphoton search gains extra competitiveness in case of a broad resonance. 

\section{Conclusion}\label{sec:conclusions}

We examined the possibility of searching for axion-like particles in central exclusive production in proton-proton collisions at the center-of-mass energy of 13 TeV for an integrated luminosity of 300 fb$^{-1}$. To quantify the exclusion power, we implemented the helicity amplitudes for light-by-light scattering induced by effective dimension-5 operators coupling the new (pseudo) scalars to photons in the publicly available Forward Physics Monte Carlo event generator. 

To mimic as best as possible the search in AFP or CT-PPS, we made a realistic offline selection on the photon pair as well as on the outgoing intact protons to better isolate quasi-elastic proton-proton collisions. A proper treatment of the background events, with the largest contribution coming from non-exclusive diphoton production overlapped with soft diffractive events, was also presented. We applied a proper statistical treatment to derive the expected bounds on the ALP--photon coupling in the mass reach covered by the exclusive channel, which is the main result of this work.

We have found that the bounds on the ALP--photon coupling for masses above 600 GeV  can be improved significantly in the central exclusive photon pair production channel. These regions are constrained by standard LHC bump searches.
In the  $0.6-2$ TeV range we expect that our exclusive diphoton search does better than existing bump searches extrapolated for $300$~fb$^{-1} $ and  $\sqrt{s}=13$~TeV  by a  factor of $\sim 3-4$ in the ALP--photon coupling. Moreover, the standard bump searches are valid only for narrow enough resonances and lose efficiency for broad resonances, which is not the case of our exclusive diphoton search method. 
 
We conclude that ALP search via light-by-light scattering in central exclusive production in p-p collisions complements other LHC searches and competes with them, especially in the  $0.6-2$ TeV mass range.

\section*{Acknowledgements}
 
 This work has been initiated at the \textit{LHC Chapter II: The Run for New Physics} workshop held at IIP.  
 SF's work is supported by the S\~ao Paulo Research Foundation (FAPESP) under grants \#2011/11973 and \#2014/21477-2. CB thanks the financial support provided by the starting research grant of CR as a Distinguished Foundation Professor.

\bibliographystyle{JHEP} 

\bibliography{references}

\providecommand{\href}[2]{#2}\begingroup\raggedright\begin{thebibliography}{10}

\bibitem{Peccei:1977hh}
R.~D. Peccei and H.~R. Quinn, \emph{{CP Conservation in the Presence of
  Instantons}}, \href{https://doi.org/10.1103/PhysRevLett.38.1440}{\emph{Phys.
  Rev. Lett.} {\bfseries 38} (1977) 1440}.

\bibitem{PhysRevLett.38.1440}
R.~D. Peccei and H.~R. Quinn, \emph{$\mathrm{CP}$ conservation in the presence
  of pseudoparticles},
  \href{https://doi.org/10.1103/PhysRevLett.38.1440}{\emph{Phys. Rev. Lett.}
  {\bfseries 38} (1977) 1440}.

\bibitem{Witten:1984dg}
E.~Witten, \emph{{Some Properties of O(32) Superstrings}},
  \href{https://doi.org/10.1016/0370-2693(84)90422-2}{\emph{Phys. Lett.}
  {\bfseries 149B} (1984) 351}.

\bibitem{Conlon:2006tq}
J.~P. Conlon, \emph{{The QCD axion and moduli stabilisation}},
  \href{https://doi.org/10.1088/1126-6708/2006/05/078}{\emph{JHEP} {\bfseries
  05} (2006) 078} [\href{https://arxiv.org/abs/hep-th/0602233}{{\ttfamily
  hep-th/0602233}}].

\bibitem{Svrcek:2006yi}
P.~Svrcek and E.~Witten, \emph{{Axions In String Theory}},
  \href{https://doi.org/10.1088/1126-6708/2006/06/051}{\emph{JHEP} {\bfseries
  06} (2006) 051} [\href{https://arxiv.org/abs/hep-th/0605206}{{\ttfamily
  hep-th/0605206}}].

\bibitem{Arvanitaki:2009fg}
A.~Arvanitaki, S.~Dimopoulos, S.~Dubovsky, N.~Kaloper and J.~March-Russell,
  \emph{{String Axiverse}},
  \href{https://doi.org/10.1103/PhysRevD.81.123530}{\emph{Phys. Rev.}
  {\bfseries D81} (2010) 123530}
  [\href{https://arxiv.org/abs/0905.4720}{{\ttfamily 0905.4720}}].

\bibitem{Acharya:2010zx}
B.~S. Acharya, K.~Bobkov and P.~Kumar, \emph{{An M Theory Solution to the
  Strong CP Problem and Constraints on the Axiverse}},
  \href{https://doi.org/10.1007/JHEP11(2010)105}{\emph{JHEP} {\bfseries 11}
  (2010) 105} [\href{https://arxiv.org/abs/1004.5138}{{\ttfamily 1004.5138}}].

\bibitem{Cicoli:2012sz}
M.~Cicoli, M.~Goodsell and A.~Ringwald, \emph{{The type IIB string axiverse and
  its low-energy phenomenology}},
  \href{https://doi.org/10.1007/JHEP10(2012)146}{\emph{JHEP} {\bfseries 10}
  (2012) 146} [\href{https://arxiv.org/abs/1206.0819}{{\ttfamily 1206.0819}}].

\bibitem{Masso:1995tw}
E.~Masso and R.~Toldra, \emph{{On a light spinless particle coupled to
  photons}}, \href{https://doi.org/10.1103/PhysRevD.52.1755}{\emph{Phys. Rev.}
  {\bfseries D52} (1995) 1755}
  [\href{https://arxiv.org/abs/hep-ph/9503293}{{\ttfamily hep-ph/9503293}}].

\bibitem{Csaki:2007ns}
C.~Csaki, J.~Hubisz and S.~J. Lee, \emph{{Radion phenomenology in realistic
  warped space models}},
  \href{https://doi.org/10.1103/PhysRevD.76.125015}{\emph{Phys. Rev.}
  {\bfseries D76} (2007) 125015}
  [\href{https://arxiv.org/abs/0705.3844}{{\ttfamily 0705.3844}}].

\bibitem{Goldberger:2008zz}
W.~D. Goldberger, B.~Grinstein and W.~Skiba, \emph{{Distinguishing the Higgs
  boson from the dilaton at the Large Hadron Collider}},
  \href{https://doi.org/10.1103/PhysRevLett.100.111802}{\emph{Phys. Rev. Lett.}
  {\bfseries 100} (2008) 111802}
  [\href{https://arxiv.org/abs/0708.1463}{{\ttfamily 0708.1463}}].

\bibitem{Fichet:2016xvs}
S.~Fichet, G.~von Gersdorff, E.~Ponton and R.~Rosenfeld, \emph{{The Excitation
  of the Global Symmetry-Breaking Vacuum in Composite Higgs Models}},
  \href{https://doi.org/10.1007/JHEP09(2016)158}{\emph{JHEP} {\bfseries 09}
  (2016) 158} [\href{https://arxiv.org/abs/1607.03125}{{\ttfamily
  1607.03125}}].

\bibitem{Bauer:2017ris}
M.~Bauer, M.~Neubert and A.~Thamm, \emph{{Collider Probes of Axion-Like
  Particles}}, \href{https://doi.org/10.1007/JHEP12(2017)044}{\emph{JHEP}
  {\bfseries 12} (2017) 044}
  [\href{https://arxiv.org/abs/1708.00443}{{\ttfamily 1708.00443}}].

\bibitem{Knapen:2016moh}
S.~Knapen, T.~Lin, H.~K. Lou and T.~Melia, \emph{{Searching for Axionlike
  Particles with Ultraperipheral Heavy-Ion Collisions}},
  \href{https://doi.org/10.1103/PhysRevLett.118.171801}{\emph{Phys. Rev. Lett.}
  {\bfseries 118} (2017) 171801}
  [\href{https://arxiv.org/abs/1607.06083}{{\ttfamily 1607.06083}}].

\bibitem{Aaboud:2017bwk}
{\scshape ATLAS} collaboration, M.~Aaboud et~al., \emph{{Evidence for
  light-by-light scattering in heavy-ion collisions with the ATLAS detector at
  the LHC}}, \href{https://doi.org/10.1038/nphys4208}{\emph{Nature Phys.}
  {\bfseries 13} (2017) 852}
  [\href{https://arxiv.org/abs/1702.01625}{{\ttfamily 1702.01625}}].

\bibitem{Knapen:2017ebd}
S.~Knapen, T.~Lin, H.~K. Lou and T.~Melia, \emph{{LHC limits on axion-like
  particles from heavy-ion collisions}},  in \emph{{Photon 2017: International
  Conference on the Structure and the Interactions of the Photon and 22th
  International Workshop on Photon-Photon Collisions and the International
  Workshop on High Energy Photon Colliders CERN, Geneva, Switzerland, May
  22-26, 2017}}, 2017, \href{https://arxiv.org/abs/1709.07110}{{\ttfamily
  1709.07110}},
  \href{https://inspirehep.net/record/1624686/files/arXiv:1709.07110.pdf}{https://inspirehep.net/record/1624686/files/arXiv:1709.07110.pdf}.

\bibitem{usww}
E.~Chapon, C.~Royon and O.~Kepka, \emph{{Anomalous quartic W W gamma gamma, Z Z
  gamma gamma, and trilinear WW gamma couplings in two-photon processes at high
  luminosity at the LHC}},
  \href{https://doi.org/10.1103/PhysRevD.81.074003}{\emph{Phys.Rev.} {\bfseries
  D81} (2010) 074003} [\href{https://arxiv.org/abs/0912.5161}{{\ttfamily
  0912.5161}}].

\bibitem{usw}
O.~Kepka and C.~Royon, \emph{{Anomalous $W W \gamma$ coupling in photon-induced
  processes using forward detectors at the LHC}},
  \href{https://doi.org/10.1103/PhysRevD.78.073005}{\emph{Phys.Rev.} {\bfseries
  D78} (2008) 073005} [\href{https://arxiv.org/abs/0808.0322}{{\ttfamily
  0808.0322}}].

\bibitem{Sahin:2009gq}
I.~Sahin and S.~C. Inan, \emph{{Probe of unparticles at the LHC in exclusive
  two lepton and two photon production via photon-photon fusion}},
  \href{https://doi.org/10.1088/1126-6708/2009/09/069}{\emph{JHEP} {\bfseries
  09} (2009) 069} [\href{https://arxiv.org/abs/0907.3290}{{\ttfamily
  0907.3290}}].

\bibitem{Atag:2010bh}
S.~Atag, S.~C. Inan and I.~Sahin, \emph{{Extra dimensions in ${\gamma\gamma
  \rightarrow \gamma\gamma}$ process at the CERN-LHC}},
  \href{https://doi.org/10.1007/JHEP09(2010)042}{\emph{JHEP} {\bfseries 09}
  (2010) 042} [\href{https://arxiv.org/abs/1005.4792}{{\ttfamily 1005.4792}}].

\bibitem{Gupta:2011be}
R.~S. Gupta, \emph{{Probing Quartic Neutral Gauge Boson Couplings using
  diffractive photon fusion at the LHC}},
  \href{https://doi.org/10.1103/PhysRevD.85.014006}{\emph{Phys.Rev.} {\bfseries
  D85} (2012) 014006} [\href{https://arxiv.org/abs/1111.3354}{{\ttfamily
  1111.3354}}].

\bibitem{Epele:2012jn}
L.~N. Epele, H.~Fanchiotti, C.~A.~G. Canal, V.~A. Mitsou and V.~Vento,
  \emph{{Looking for magnetic monopoles at LHC with diphoton events}},
  \href{https://doi.org/10.1140/epjp/i2012-12060-8}{\emph{Eur. Phys. J. Plus}
  {\bfseries 127} (2012) 60} [\href{https://arxiv.org/abs/1205.6120}{{\ttfamily
  1205.6120}}].

\bibitem{Lebiedowicz:2013fta}
P.~Lebiedowicz, R.~Pasechnik and A.~Szczurek, \emph{{Search for technipions in
  exclusive production of diphotons with large invariant masses at the LHC}},
  \href{https://doi.org/10.1016/j.nuclphysb.2014.02.008}{\emph{Nucl. Phys.}
  {\bfseries B881} (2014) 288}
  [\href{https://arxiv.org/abs/1309.7300}{{\ttfamily 1309.7300}}].

\bibitem{Fichet:2013ola}
S.~Fichet and G.~von Gersdorff, \emph{{Anomalous gauge couplings from composite
  Higgs and warped extra dimensions}}, {\emph{JHEP03(2014)102} (2013) }
  [\href{https://arxiv.org/abs/1311.6815}{{\ttfamily 1311.6815}}].

\bibitem{Fichet:2013gsa}
S.~Fichet, G.~von Gersdorff, O.~Kepka, B.~Lenzi, C.~Royon and M.~Saimpert,
  \emph{{Probing new physics in diphoton production with proton tagging at the
  Large Hadron Collider}},
  \href{https://doi.org/10.1103/PhysRevD.89.114004}{\emph{Phys. Rev.}
  {\bfseries D89} (2014) 114004}
  [\href{https://arxiv.org/abs/1312.5153}{{\ttfamily 1312.5153}}].

\bibitem{Sun:2014qoa}
H.~Sun, \emph{{Probe anomalous $tq\gamma$ couplings through single top
  photoproduction at the LHC}},
  \href{https://doi.org/10.1016/j.nuclphysb.2014.07.012}{\emph{Nucl.Phys.}
  {\bfseries B886} (2014) 691}
  [\href{https://arxiv.org/abs/1402.1817}{{\ttfamily 1402.1817}}].

\bibitem{Sun:2014qba}
H.~Sun, \emph{{Large Extra Dimension effects through Light-by-Light Scattering
  at the CERN LHC}},
  \href{https://doi.org/10.1140/epjc/s10052-014-2977-1}{\emph{Eur.Phys.J.}
  {\bfseries C74} (2014) 2977}
  [\href{https://arxiv.org/abs/1406.3897}{{\ttfamily 1406.3897}}].

\bibitem{Sun:2014ppa}
H.~Sun, \emph{{Dark Matter Searches in Jet plus Missing Energy in $\rm \gamma
  p$ collision at CERN LHC}},
  \href{https://doi.org/10.1103/PhysRevD.90.035018}{\emph{Phys.Rev.} {\bfseries
  D90} (2014) 035018} [\href{https://arxiv.org/abs/1407.5356}{{\ttfamily
  1407.5356}}].

\bibitem{Sahin:2014dua}
I.~Sahin, M.~Koksal, S.~C. Inan, A.~A. Billur, B.~Sahin, P.~Tektas et~al.,
  \emph{{Graviton production through photon-quark scattering at the LHC}},
  \href{https://doi.org/10.1103/PhysRevD.91.035017}{\emph{Phys. Rev.}
  {\bfseries D91} (2015) 035017}
  [\href{https://arxiv.org/abs/1409.1796}{{\ttfamily 1409.1796}}].

\bibitem{Inan:2014mua}
S.~C. Inan, \emph{{Dimension-six anomalous $tq\gamma$ couplings in
  $\gamma\gamma$ collision at the LHC}},
  \href{https://doi.org/10.1016/j.nuclphysb.2015.05.028}{\emph{Nucl. Phys.}
  {\bfseries B897} (2015) 289}
  [\href{https://arxiv.org/abs/1410.3609}{{\ttfamily 1410.3609}}].

\bibitem{Fichet:2014uka}
S.~Fichet, G.~von Gersdorff, B.~Lenzi, C.~Royon and M.~Saimpert,
  \emph{{Light-by-light scattering with intact protons at the LHC: from
  Standard Model to New Physics}},
  \href{https://doi.org/10.1007/JHEP02(2015)165}{\emph{JHEP} {\bfseries 02}
  (2015) 165} [\href{https://arxiv.org/abs/1411.6629}{{\ttfamily 1411.6629}}].

\bibitem{Fichet:2015nia}
S.~Fichet, \emph{{Prospects for new physics searches at the LHC in the forward
  proton mode}}, \href{https://doi.org/10.5506/APhysPolBSupp.8.811}{\emph{Acta
  Phys. Polon. Supp.} {\bfseries 8} (2015) 811}
  [\href{https://arxiv.org/abs/1510.01004}{{\ttfamily 1510.01004}}].

\bibitem{Cho:2015dha}
G.-C. Cho, T.~Kono, K.~Mawatari and K.~Yamashita, \emph{{Search for
  Kaluza-Klein gravitons in extra dimension models via forward detectors at the
  LHC}}, \href{https://doi.org/10.1103/PhysRevD.91.115015}{\emph{Phys. Rev.}
  {\bfseries D91} (2015) 115015}
  [\href{https://arxiv.org/abs/1503.05678}{{\ttfamily 1503.05678}}].

\bibitem{Fichet:2016clq}
S.~Fichet, \emph{{Shining Light on Polarizable Dark Particles}},
  \href{https://doi.org/10.1007/JHEP04(2017)088}{\emph{JHEP} {\bfseries 04}
  (2017) 088} [\href{https://arxiv.org/abs/1609.01762}{{\ttfamily
  1609.01762}}].

\bibitem{Fichet:2015vvy}
S.~Fichet, G.~von Gersdorff and C.~Royon, \emph{{Scattering light by light at
  750 GeV at the LHC}},
  \href{https://doi.org/10.1103/PhysRevD.93.075031}{\emph{Phys. Rev.}
  {\bfseries D93} (2016) 075031}
  [\href{https://arxiv.org/abs/1512.05751}{{\ttfamily 1512.05751}}].

\bibitem{Fichet:2016pvq}
S.~Fichet, G.~von Gersdorff and C.~Royon, \emph{{Measuring the Diphoton
  Coupling of a 750 GeV Resonance}},
  \href{https://doi.org/10.1103/PhysRevLett.116.231801}{\emph{Phys. Rev. Lett.}
  {\bfseries 116} (2016) 231801}
  [\href{https://arxiv.org/abs/1601.01712}{{\ttfamily 1601.01712}}].

\bibitem{Baldenegro:2017aen}
C.~Baldenegro, S.~Fichet, G.~von Gersdorff and C.~Royon, \emph{{Probing the
  anomalous γγγZ coupling at the LHC with proton tagging}},
  \href{https://doi.org/10.1007/JHEP06(2017)142}{\emph{JHEP} {\bfseries 06}
  (2017) 142} [\href{https://arxiv.org/abs/1703.10600}{{\ttfamily
  1703.10600}}].

\bibitem{Baldenegro:2017lzv}
S.~Fichet and C.~Baldenegro, \emph{{Anomalous gauge interactions in photon
  collisions at the LHC and the FCC}},  in \emph{{Photon 2017, CERN, Geneva,
  Switzerland, May 22-26, 2017}}, 2017,
  \href{https://arxiv.org/abs/1708.07531}{{\ttfamily 1708.07531}},
  \href{http://inspirehep.net/record/1620061/files/arXiv:1708.07531.pdf}{http://inspirehep.net/record/1620061/files/arXiv:1708.07531.pdf}.

\bibitem{Adamczyk:2017378}
L.~Adamczyk, E.~Banaś, A.~Brandt, M.~Bruschi, S.~Grinstein, J.~Lange et~al.,
  \emph{{Technical Design Report for the ATLAS Forward Proton Detector}},
  Tech. Rep. CERN-LHCC-2015-009. ATLAS-TDR-024, May, 2015.

\bibitem{Albrow:1753795}
M.~Albrow, M.~Arneodo, V.~Avati, J.~Baechler, N.~Cartiglia, M.~Deile et~al.,
  \emph{{CMS-TOTEM Precision Proton Spectrometer}},  Tech. Rep.
  CERN-LHCC-2014-021. TOTEM-TDR-003. CMS-TDR-13, Sep, 2014.

\bibitem{Cms:2018het}
{\scshape TOTEM} collaboration, Cms, \emph{{Observation of proton-tagged,
  central (semi)exclusive production of high-mass lepton pairs in pp collisions
  at 13 TeV with the CMS-TOTEM precision proton spectrometer}},
  \href{https://arxiv.org/abs/1803.04496}{{\ttfamily 1803.04496}}.

\bibitem{Terazawa:1973tb}
H.~Terazawa, \emph{{Two photon processes for particle production at
  high-energies}},
  \href{https://doi.org/10.1103/RevModPhys.45.615}{\emph{Rev.Mod.Phys.}
  {\bfseries 45} (1973) 615}.

\bibitem{Budnev}
V.~Budnev, I.~Ginzburg, G.~Meledin and V.~Serbo, \emph{{The Two photon particle
  production mechanism. Physical problems. Applications. Equivalent photon
  approximation}},
  \href{https://doi.org/10.1016/0370-1573(75)90009-5}{\emph{Phys.Rept.}
  {\bfseries 15} (1975) 181}.

\bibitem{SuperChic2}
L.~A. Harland-Lang, V.~A. Khoze and M.~G. Ryskin, \emph{{Exclusive physics at
  the LHC with SuperChic 2}},
  \href{https://doi.org/10.1140/epjc/s10052-015-3832-8}{\emph{Eur. Phys. J.}
  {\bfseries C76} (2016) 9} [\href{https://arxiv.org/abs/1508.02718}{{\ttfamily
  1508.02718}}].

\bibitem{Khoze:2017sdd}
V.~A. Khoze, A.~D. Martin and M.~G. Ryskin, \emph{{Multiple interactions and
  rapidity gap survival}},  \href{https://arxiv.org/abs/1710.11505}{{\ttfamily
  1710.11505}}.

\bibitem{Khoze:2002dc}
V.~A. Khoze, A.~D. Martin and M.~G. Ryskin, \emph{{Photon exchange processes at
  hadron colliders as a probe of the dynamics of diffraction}},
  \href{https://doi.org/10.1007/s10052-002-0964-4}{\emph{Eur. Phys. J.}
  {\bfseries C24} (2002) 459}
  [\href{https://arxiv.org/abs/hep-ph/0201301}{{\ttfamily hep-ph/0201301}}].

\bibitem{Chatrchyan:2012tv}
{\scshape CMS} collaboration, S.~Chatrchyan et~al., \emph{{Search for exclusive
  or semi-exclusive photon pair production and observation of exclusive and
  semi-exclusive electron pair production in $pp$ collisions at $\sqrt{s}=7$
  TeV}}, \href{https://doi.org/10.1007/JHEP11(2012)080}{\emph{JHEP} {\bfseries
  11} (2012) 080} [\href{https://arxiv.org/abs/1209.1666}{{\ttfamily
  1209.1666}}].

\bibitem{Khachatryan:2015iwa}
{\scshape CMS} collaboration, V.~Khachatryan et~al., \emph{{Performance of
  Photon Reconstruction and Identification with the CMS Detector in
  Proton-Proton Collisions at sqrt(s) = 8 TeV}},
  \href{https://doi.org/10.1088/1748-0221/10/08/P08010}{\emph{JINST} {\bfseries
  10} (2015) P08010} [\href{https://arxiv.org/abs/1502.02702}{{\ttfamily
  1502.02702}}].

\bibitem{MITREVSKI20162539}
J.~Mitrevski, \emph{Electron and photon reconstruction with the atlas
  detector},
  \href{https://doi.org/https://doi.org/10.1016/j.nuclphysbps.2015.09.452}{\emph{Nuclear
  and Particle Physics Proceedings} {\bfseries 273-275} (2016) 2539 }.

\bibitem{Khoze:2001xm}
V.~Khoze, A.~Martin and M.~Ryskin, \emph{{Prospects for new physics
  observations in diffractive processes at the LHC and Tevatron}},
  \href{https://doi.org/10.1007/s100520100884}{\emph{Eur.Phys.J.} {\bfseries
  C23} (2002) 311} [\href{https://arxiv.org/abs/hep-ph/0111078}{{\ttfamily
  hep-ph/0111078}}].

\bibitem{FPMC}
M.~Boonekamp, A.~Dechambre, V.~Juranek, O.~Kepka, M.~Rangel et~al.,
  \emph{{FPMC: A Generator for forward physics}},
  \href{https://arxiv.org/abs/1102.2531}{{\ttfamily 1102.2531}}.

\bibitem{Royon:2006by}
C.~Royon, L.~Schoeffel, S.~Sapeta, R.~B. Peschanski and E.~Sauvan, \emph{{A
  Global analysis of inclusive diffractive cross sections at HERA}},
  \href{https://doi.org/10.1016/j.nuclphysb.2007.05.016}{\emph{Nucl. Phys.}
  {\bfseries B781} (2007) 1}
  [\href{https://arxiv.org/abs/hep-ph/0609291}{{\ttfamily hep-ph/0609291}}].

\bibitem{INGELMAN1985256}
G.~Ingelman and P.~Schlein, \emph{Jet structure in high mass diffractive
  scattering},
  \href{https://doi.org/https://doi.org/10.1016/0370-2693(85)91181-5}{\emph{Physics
  Letters B} {\bfseries 152} (1985) 256 }.

\bibitem{herwig}
G.~Corcella, I.~Knowles, G.~Marchesini, S.~Moretti, K.~Odagiri et~al.,
  \emph{{HERWIG 6.5 release note}},
  \href{https://arxiv.org/abs/hep-ph/0210213}{{\ttfamily hep-ph/0210213}}.

\bibitem{survival1}
V.~Khoze, A.~Martin and M.~Ryskin, \emph{{Prospects for new physics
  observations in diffractive processes at the LHC and Tevatron}},
  \href{https://doi.org/10.1007/s100520100884}{\emph{Eur.Phys.J.} {\bfseries
  C23} (2002) 311} [\href{https://arxiv.org/abs/hep-ph/0111078}{{\ttfamily
  hep-ph/0111078}}].

\bibitem{survival2}
E.~Gotsman, E.~Levin, U.~Maor, E.~Naftali and A.~Prygarin, \emph{{Survival
  probability of large rapidity gaps}},
  \href{https://arxiv.org/abs/hep-ph/0511060}{{\ttfamily hep-ph/0511060}}.

\bibitem{Sjostrand:2007gs}
T.~Sjostrand, S.~Mrenna and P.~Z. Skands, \emph{{A Brief Introduction to PYTHIA
  8.1}}, \href{https://doi.org/10.1016/j.cpc.2008.01.036}{\emph{Comput. Phys.
  Commun.} {\bfseries 178} (2008) 852}
  [\href{https://arxiv.org/abs/0710.3820}{{\ttfamily 0710.3820}}].

\bibitem{Fastjet}
M.~Cacciari, G.~P. Salam and G.~Soyez, \emph{{FastJet User Manual}},
  \href{https://doi.org/10.1140/epjc/s10052-012-1896-2}{\emph{Eur. Phys. J.}
  {\bfseries C72} (2012) 1896}
  [\href{https://arxiv.org/abs/1111.6097}{{\ttfamily 1111.6097}}].

\bibitem{Timing}
C.~Royon, M.~Saimpert, O.~Kepka and R.~Zlebcik, \emph{{Timing detectors for
  proton tagging at the LHC}}, {\emph{{Acta Physica Polonica B Proceedings
  supplement}} {\bfseries 7} (2014) 735}.

\bibitem{Ferreira:2017ymn}
F.~Ferreira, S.~Fichet and V.~Sanz, \emph{{On new physics searches with
  multidimensional differential shapes}},
  \href{https://doi.org/10.1016/j.physletb.2018.01.008}{\emph{Phys. Lett.}
  {\bfseries B778} (2018) 35}
  [\href{https://arxiv.org/abs/1702.05106}{{\ttfamily 1702.05106}}].

\bibitem{PhysRevLett.59.755}
E.~M. Riordan, M.~W. Krasny, K.~Lang, P.~de~Barbaro, A.~Bodek, S.~Dasu et~al.,
  \emph{Search for short-lived axions in an electron-beam-dump experiment},
  \href{https://doi.org/10.1103/PhysRevLett.59.755}{\emph{Phys. Rev. Lett.}
  {\bfseries 59} (1987) 755}.

\bibitem{PhysRevD.38.3375}
J.~D. Bjorken, S.~Ecklund, W.~R. Nelson, A.~Abashian, C.~Church, B.~Lu et~al.,
  \emph{Search for neutral metastable penetrating particles produced in the
  slac beam dump}, \href{https://doi.org/10.1103/PhysRevD.38.3375}{\emph{Phys.
  Rev. D} {\bfseries 38} (1988) 3375}.

\bibitem{Döbrich2016}
B.~D{\"o}brich, J.~Jaeckel, F.~Kahlhoefer, A.~Ringwald and K.~Schmidt-Hoberg,
  \emph{Alptraum: Alp production in proton beam dump experiments},
  {\emph{Journal of High Energy Physics} {\bfseries 2016} (2016) 18}.

\bibitem{PhysRevD.51.2053}
R.~Balest et~al.,
  \emph{\ensuremath{\Upsilon}(1s)\ensuremath{\rightarrow}\ensuremath{\gamma}+noninteracting
  particles}, \href{https://doi.org/10.1103/PhysRevD.51.2053}{\emph{Phys. Rev.
  D} {\bfseries 51} (1995) 2053}.

\bibitem{delAmoSanchez:2010ac}
{\scshape BaBar} collaboration, P.~del Amo~Sanchez et~al., \emph{{Search for
  Production of Invisible Final States in Single-Photon Decays of
  $\Upsilon(1S)$}},
  \href{https://doi.org/10.1103/PhysRevLett.107.021804}{\emph{Phys. Rev. Lett.}
  {\bfseries 107} (2011) 021804}
  [\href{https://arxiv.org/abs/1007.4646}{{\ttfamily 1007.4646}}].

\bibitem{Jaeckel:2012yz}
J.~Jaeckel, M.~Jankowiak and M.~Spannowsky, \emph{{LHC probes the hidden
  sector}}, \href{https://doi.org/10.1016/j.dark.2013.06.001}{\emph{Phys. Dark
  Univ.} {\bfseries 2} (2013) 111}
  [\href{https://arxiv.org/abs/1212.3620}{{\ttfamily 1212.3620}}].

\bibitem{Mimasu:2014nea}
K.~Mimasu and V.~Sanz, \emph{{ALPs at Colliders}},
  \href{https://doi.org/10.1007/JHEP06(2015)173}{\emph{JHEP} {\bfseries 06}
  (2015) 173} [\href{https://arxiv.org/abs/1409.4792}{{\ttfamily 1409.4792}}].

\bibitem{Jaeckel:2015jla}
J.~Jaeckel and M.~Spannowsky, \emph{{Probing MeV to 90 GeV axion-like particles
  with LEP and LHC}},
  \href{https://doi.org/10.1016/j.physletb.2015.12.037}{\emph{Phys. Lett.}
  {\bfseries B753} (2016) 482}
  [\href{https://arxiv.org/abs/1509.00476}{{\ttfamily 1509.00476}}].

\end{thebibliography}\endgroup

\end{document}